\documentclass[12pt]{iopart}
\usepackage{iopams}  
\usepackage{hyperref}
\usepackage{color}
\usepackage[a4paper]{anysize}

\definecolor{darkred}{rgb}{0.5,0.0,0.0}
\definecolor{darkgreen}{rgb}{0.0,0.5,0.0}
\definecolor{darkblue}{rgb}{0.0,0.0,0.5}
\hypersetup{colorlinks=true,
citecolor=darkred,
linkcolor=darkgreen,
urlcolor=darkblue}
\marginsize{3cm}{2.5cm}{2.0cm}{4.0cm}

\usepackage{graphicx}
\usepackage{bm}
\usepackage{bbm} 
\usepackage[applemac]{inputenc}

\begin{document} 
  
\title[MAR transport and critical current in TS nanowire junctions]{Multiple Andreev reflection and critical current in topological superconducting nanowire junctions}
\author{Pablo San-Jose$^1$, Jorge Cayao$^1$, Elsa Prada$^2$, Ramón Aguado$^1$}
\address{$^1$Instituto de Ciencia de Materiales de Madrid (ICMM), Consejo Superior de Investigaciones Científicas (CSIC), Sor Juana Inés de la Cruz 3, 28049 Madrid, Spain\\$^2$Departamento de Física de la Materia Condensada, Universidad Autónoma de Madrid, Cantoblanco, 28049 Madrid, Spain}

\ead{pablo.sanjose@csic.es}


\begin{abstract}
We study transport in a voltage biased superconductor-normal-superconductor (SNS) junction made of semiconducting nanowires with strong spin-orbit coupling, as it transitions into a topological superconducting phase for increasing Zeeman field. Despite the absence of a fractional steady-state ac Josephson current in the topological phase, the dissipative multiple Andreev reflection (MAR) current $I_{dc}$ at different junction transparencies is particularly revealing. It exhibits unique  features related to topology, such as the gap inversion, the formation of Majorana bound states, and fermion-parity conservation. Moreover, the critical current $I_c$, which remarkably does not vanish at the critical point where the system becomes gapless, provides direct evidence of the topological transition. 
\end{abstract}

\maketitle
\section{Introduction}
Semiconducting nanowires (NWs) with a strong spin-orbit (SO) coupling in the proximity of s-wave superconductors and in the presence of an external Zeeman magnetic field $B$ are a promising platform to study Majorana physics. Theory predicts that above a critical field $B_c\equiv\sqrt{\mu^2+\Delta^2}$, defined in terms of the Fermi energy $\mu$ and the induced s-wave pairing $\Delta$, the wire undergoes a topological transition into a phase hosting zero energy Majorana bound states (MBSs) at the ends of the wire \cite{Lutchyn:PRL10,Oreg:PRL10}. Recent experiments have reported measurements of differential conductance $dI/dV$ that support the existence of such MBSs at normal-superconductor (NS) junctions in InSb  \cite{Delft-exp,Deng:NL12} and InAs \cite{Das:NP12} NWs.
The main result of these experiments is an emergent zero-bias anomaly (ZBA) in $dI/dV$ as $B$ increases. 
In this context, the ZBA results from  tunnelling into the MBS \cite{Sengupta:PRB01,Bolech:PRL07,Law:PRL09}.
Although these experiments are partially consistent with the MBS interpretation \cite{Prada:PRB12,Stanescu:PRL12,Rainis:PRB13,Pientka:PRL12}, some important features such as the expected superconducting gap inversion were not observed. Moreover, other mechanisms that give rise to ZBAs, such as disorder \cite{Pientka:PRL12,Bagrets:PRL12,Liu:PRL12,Sau:13}, Kondo physics \cite{Lee:PRL12}, or Andreev bound states  (ABSs) \cite{Finck:PRL13,Lee:13}, cannot be completely ruled out. In particular Zeeman-resolved ABSs in nanowires with charging effects  \cite{Lee:13} can give magnetic field dependencies essentially undistinguishable from some of the claimed Majorana experiments \cite{Das:NP12}. Furthermore, it has been recently pointed out that even ZBAs similar to the 0.7 anomaly in quantum point contacts may play a role in single barrier structures \cite{Churchill:PRB13}. 

Stronger evidence could be provided by the observation of non-Abelian interference (braiding) \cite{Nayak:RMP08}, or by transport in phase-sensitive superconductor-normal-superconductor (SNS) junctions. The latter approach, which typically involves the measurement of an anomalous ``fractional" $4\pi$-periodic ac Josephson effect \cite{Kitaev:P01,Fu:PRB09,Kwon:EPJB03}, is much less demanding than performing braiding. Realistically, however, the fractional effect, detected through, e.g., the absence of odd steps in Shapiro experiments \cite{Kwon:EPJB03,Jiang:PRL11,Dominguez:PRB12,Rokhinson:NP12}, may be difficult to measure (dissipation is expected to destroy it in the steady state), or may even develop without relation to topology \cite{Sau:12b}. 
Although it has been shown that the $4\pi$ periodicity survives in the dynamics, such as noise and transients \cite{Badiane:PRL11, San-Jose:PRL12a,Pikulin:PRB12}, simpler experimental probes of MBSs are extremely desirable.

Here we propose the multiple Andreev reflection (MAR) current in voltage-biased SNS junctions made of NWs \cite{Doh:S05,Nilsson:NL12}
as an alternative, remarkably powerful, yet simple tool to study the topological transition. 
This is made possible by the direct effect that gap inversion, MBS formation and fermion-parity conservation have on the MAR current $I_{dc}(V)$ at various junction transparencies $T_N$. 
For tunnel junctions, $I_{dc}(V)$ traces the closing and reopening of the superconducting gap at $B_c$, $\Delta_\mathrm{eff}\sim|B-B_c|$. This gap inversion can be shown to be a true topological transition by tuning the junction to perfect transparency $T_N=1$. In this regime, the limiting current $I_{dc}(V\to 0)$ shows signatures of the parity conservation effects that are responsible for the fractional Josephson current in the presence of MBSs, but which, in contrast to the latter, survive in the steady state limit. Moreover, the detailed dependence of MAR as a function of $T_N$ has the fundamental advantage over NS junctions in that it contains information about the peculiar dependence of MBS hybridization with superconductor phase difference $\phi$, despite not requiring any external control on it. Similarly, we show that another important phase-insensitive quantity, the critical current $I_c$, remains unexpectedly finite for all $B$ due to a significant continuum contribution, and exhibits an anomaly at the topological transition.

This paper is organized as follows. In section \ref{Rashba NW}, we review the model for Rashba nanowires in the presence of both s-wave superconducting pairing and an external Zeeman field and describe how a
spinless p-wave superconductor regime can be achieved. In particular, we discuss how the problem can be understood in terms of two independent p-wave superconductors, originated from the Rashba helical bands, and weakly coupled by an interband pairing term. This two-band description is very useful in order to understand the main results of this paper. Such results are discussed in section \ref{SNS} which is divided in two parts. The first part (subsection  \ref{ABS-SNS}) is devoted to the ABSs which are confined in the junction. The detailed evolution of these ABSs as the system undergoes a topological transition has not been discussed in the literature, to the best of our knowledge, and becomes essential in order to gain a deep understanding of transport across the junction, which is discussed in subsection \ref{ac Josephson}. In this part we study the ac Josephson effect in nanowire SNS junctions, with focus on how the MAR currents reflect the topological transition in the nanowires as the Zeeman field increases. In particular, we present a thorough analysis of MAR transport in topological SNS junctions for arbitrary transparency of the normal part. Finally, we discuss in section \ref{Critical current} how the critical current $I_c$ does not vanish at the critical point where the system becomes gapless and, importantly, how $I_c$ provides direct evidence of the topological transition. 
%

\begin{figure} 
   \includegraphics[width=1 \columnwidth]{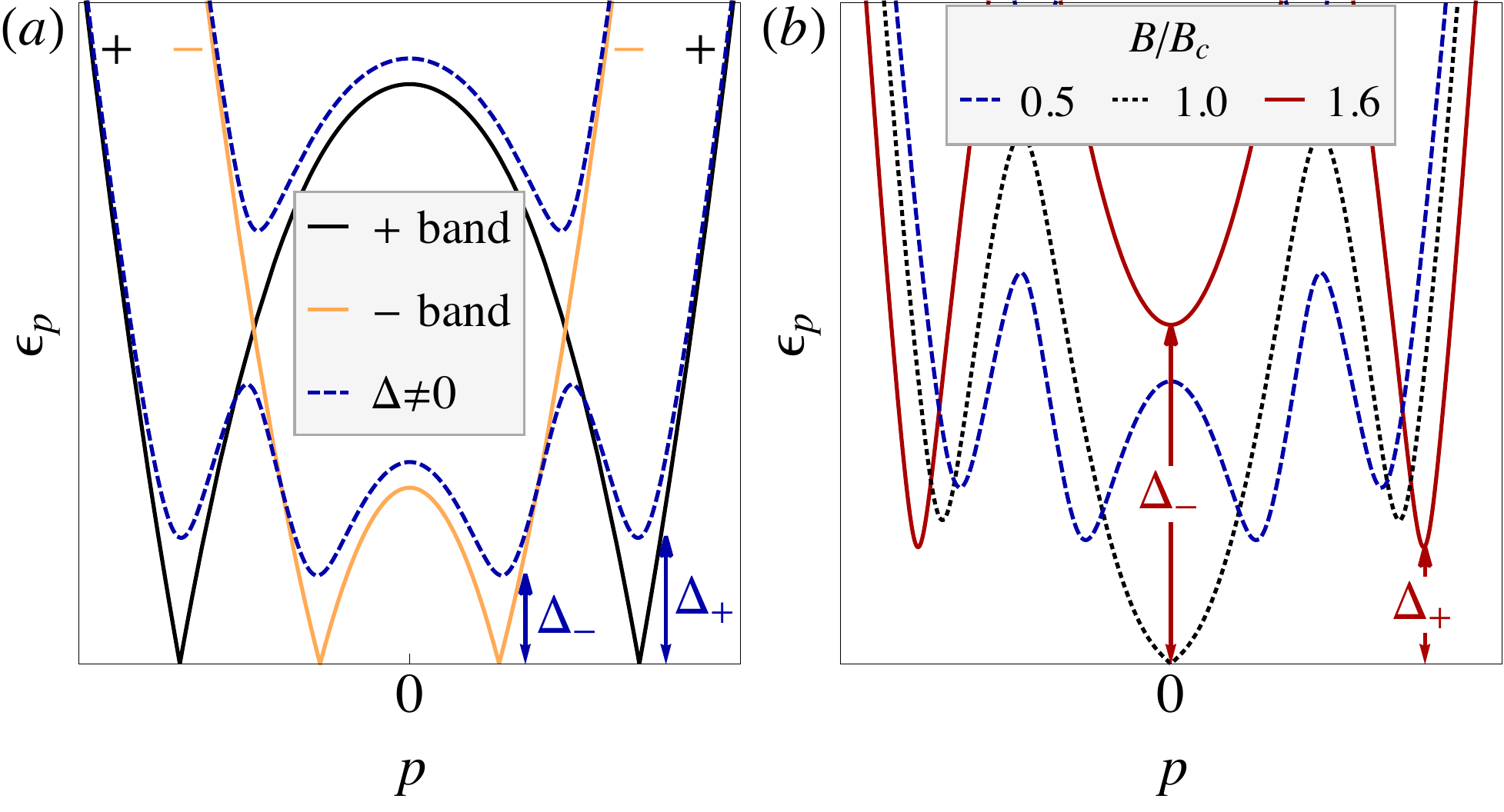} 
     \caption{(Color online) (a) Lowest bands of a $B=0.5B_c$ nanowire, with (dashed) and without (solid) pairing $\Delta$. (b) Evolution of bands with Zeeman field $B$. Gap $\Delta_-$ closes at $B=B_c$, while $\Delta_+$ does not.
}
   \label{fig:bands}
\end{figure}

\section{Rashba nanowire model and effective p-wave pairing}\label{Rashba NW}
A single one-dimensional NW in the normal state is described by the Hamiltonian  
\[
H_0=\frac{p^2}{2m^*}-\mu+\alpha_\mathrm{so}\sigma_y p+B\sigma_x,
\]
where $m^*$ is the effective mass, $\alpha_\mathrm{so}$ the SO coupling, $\mu$ the Fermi energy and $\sigma_i$ the spin Pauli matrices. An external magnetic field $\mathcal{B}$ along the wire produces a Zeeman splitting $B=\frac{1}{2}g\mu_B \mathcal{B}$, where $\mu_B$ is the Bohr
magneton and $g$ is the wire g-factor. 
The Nambu Hamiltonian
\begin{equation}\label{model}
H=\left[\begin{array}{cc}
H_0 & -i\Delta\sigma_y\\
i\Delta^*\sigma_y & -H^*_0\\
\end{array}\right],
\end{equation}
models the NW in the presence of an induced s-wave superconducting pairing $\Delta$ (here assumed real without loss of generality).
The essential ingredient for a topological superconductor is an effective p-wave pairing acting on a single (``spinless") fermionic species \cite{Kitaev:P01}. SO coupling splits NW states into two subbands of opposite helicity at $B=0$. At finite $B$, these two subbands, which we label $+$ and $-$ [black and orange lines in Fig. \ref{fig:bands}(a)], have spins canted away from the SO axis. The s-wave pairing $\Delta$, expressed in the $\pm$ basis, takes the form of an intraband p-wave $\Delta^{++/--}_{p}(p)= \pm i p\Delta\alpha_\mathrm{so}/\sqrt{B^2+(\alpha_\mathrm{so} p)^2}$, plus an interband s-wave pairing $\Delta^{+-}_s(p)=\Delta B/\sqrt{B^2+(\alpha_\mathrm{so} p)^2}$ \cite{Alicea:RPP12}. 
Without the latter, the problem decouples into two independent p-wave superconductors, while $\Delta^{+-}_s$ acts as a weak coupling between them. Each quasi-independent $\pm$ sector has a different ($B$-dependent) gap, which we call $\Delta_-$ (at small $p$) and $\Delta_+$ (large $p$), see Fig. \ref{fig:bands}(a,b). While $\Delta_+$ remains roughly constant with $B$ (for strong SO coupling  \cite{Prada:PRB12,Rainis:PRB13}), $\Delta_-$ vanishes linearly as $B$ approaches the critical field, $\Delta_-\approx |B-B_c|$ \footnote{Note that, in general, $\Delta_-$ is at a small but finite momentum. However, as $B$ approaches $B_c$, $\Delta_-$ becomes centered at $p=0$ and is approximately equal to $|E_0|$, where $E_0$ is the zero momentum energy of the lowest subband, $E_0=B-B_c$, and is related to the topological charge of the lowest superconducting band. \cite{Ghosh:PRB10}}. This closing and reopening (gap inversion) signals a topological transition, induced by the effective removal of the $-$ sector away from the low-energy problem. Below $B_c$ the NW is composed of two 
spinless p-wave superconductors, and is therefore topologically trivial. Above $B_c$, $\Delta_-$ is no longer a p-wave gap, but rather a normal (Zeeman) spectral gap already present in the normal state, transforming the wire into a single-species p-wave superconductor with non-trivial topology. This phase contains MBSs, protected by the effective gap $\Delta_\mathrm{eff}=\mathrm{Min}(\Delta_+,\Delta_-)$, at the wire ends. Above a certain field $B_c^{(2)}$, the gap $\Delta_\mathrm{eff}$ saturates at $\Delta_+$ and the physics of superconducting helical edge states in spin-Hall insulators is recovered \cite{Fu:PRB09,Badiane:PRL11}.

\section{Nanowire SNS junctions}\label{SNS}

\begin{figure}[t] 
   \centering
   \includegraphics[width=0.5\columnwidth]{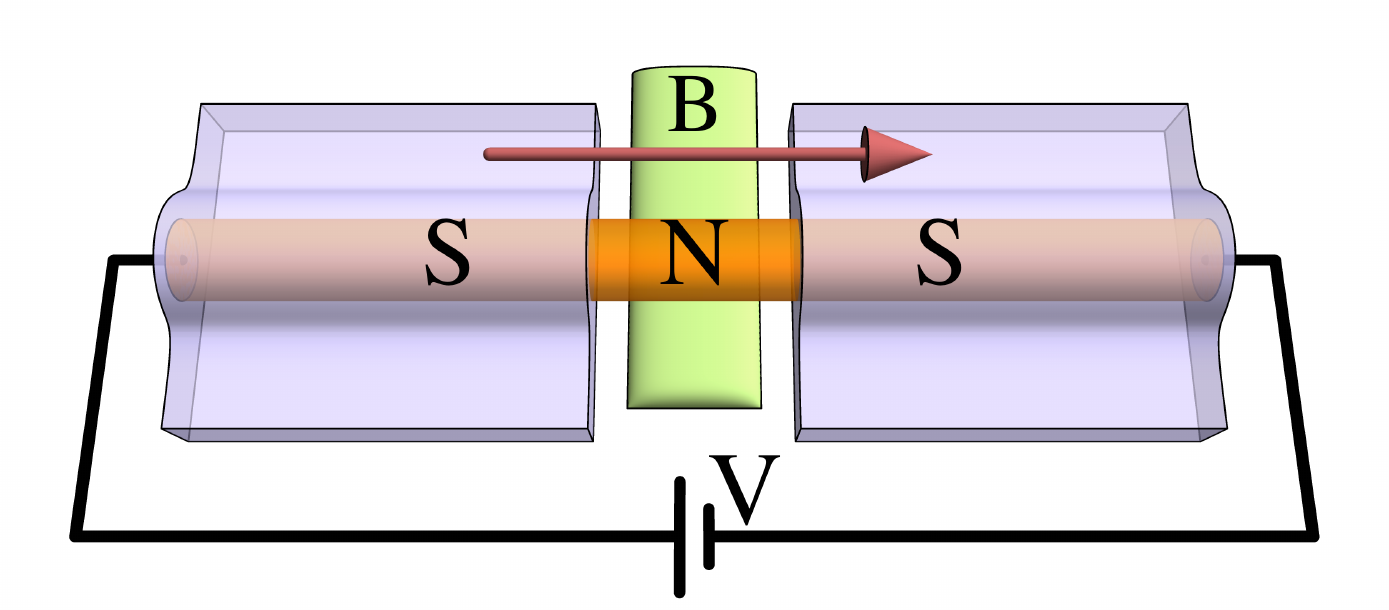}
     \caption{(Color online)      Short SNS junction fabricated by covering a semiconductor nanowire with two S-wave superconductors. A bias $V$ and a longitudinal Zeeman field $B$ can be applied to the wire. The central normal region has tuneable transparency via a depletion bottom gate.
}
   \label{fig:sys}
\end{figure}
In the previous section we described how a semiconducting nanowire with a strong SO coupling in the proximity of an s-wave superconductor and in the presence of an external Zeeman magnetic field $B$ behaves as a topological superconductor above a critical field $B_c$.
Here we are concerned with the effects of this topological transition on the MAR current $I_{dc}(V)$ across junctions formed with such nanowires. In particular, we consider SNS junctions of different normal transparencies $T_N$. Experimentally, such geometry can be fabricated by partially covering a single NW with two superconducting leads and leaving an uncovered normal region in the middle. The coupling of the normal part of the NW to the superconducting leads can be tuned by local control of the electron density with a gated constriction. This can be realized by using, e.g., bottom-gates forming a quantum point contact, see Fig. \ref{fig:sys}. Such geometry has been successfully implemented experimentally in Ref.  \cite{Churchill:PRB13} for NS junctions, where control of the coupling between the superconducting and normal sections from near pinch-off (tunneling limit) to the multichannel regime is demonstrated.
\begin{figure}[t] 
   \centering
   \includegraphics[width=\columnwidth]{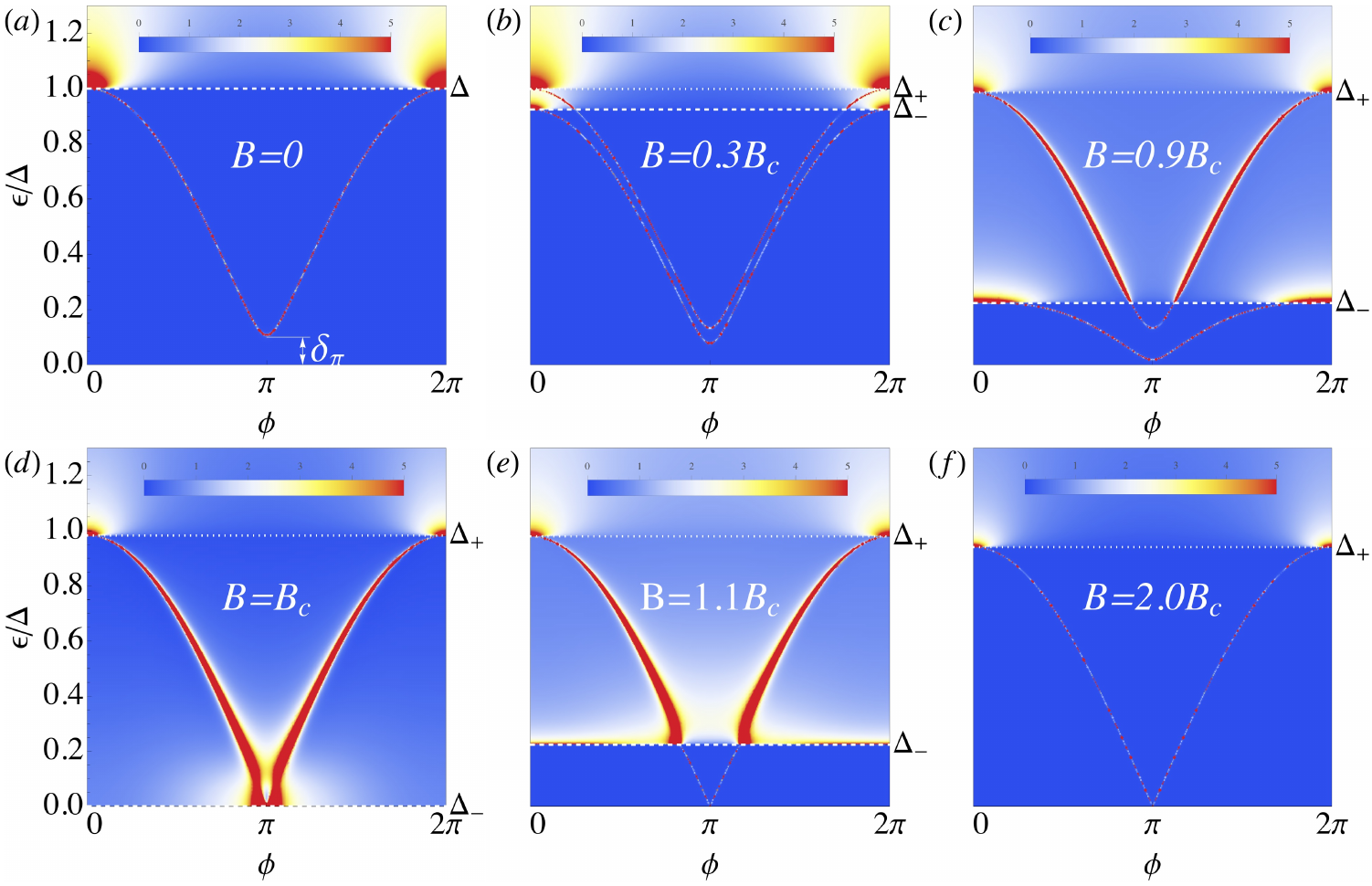} 
     \caption{(Color online)      Local density of states at the junction for perfect normal transparency $T_N=1$, which is peaked at the energy $\epsilon_\pm(\phi)$ of Andreev (quasi)bound states. Different panels show how the Andreev states evolve as the system undergoes the topological transition.
}
   \label{fig:ABS}
\end{figure}
For simplicity, here we focus on short SNS junctions\footnote{Results for the long junction limit will be published elsewhere \cite{Cayao:UP}.} with single channel nanowires. For computation purposes, we consider a discretisation of the continuum model Eq. (\ref{model}) for the Rashba nanowire into a tight-binding lattice with a small lattice spacing $a$. This transforms terms containing the momentum operator $p$ into nearest-neighbour hopping matrices $v$. Namely $H_0=\sum_{i}c^+_ihc_i+\sum_{\langle ij\rangle}c^+_i vc_j + \mathrm{h.c.}$,  with
\[
h=\left(\begin{array}{cc}2t-\mu & B \\B & 2t-\mu\end{array}\right),\hspace{0.5cm} v=\left(\begin{array}{cc}-t & \frac{\hbar}{2a}\alpha_\mathrm{so} \\-\frac{\hbar}{2a}\alpha_\mathrm{so} & -t\end{array}\right),
\] 
are matrices in spin space, and $t=\hbar^2/2m^*a^2$. The pairing is incorporated like in Eq. (\ref{model}). A short SNS junction is modelled by suppressing the hopping matrix $v_0=\nu v$ between two sites in the middle of the wire, which represent the junction. The  dimensionless factor $\nu\in[0,1]$ controls the junction's normal transparency at $B=0$, which we denote $T_N(\nu)$. A phase difference $\phi$ across the junction is implemented by multiplying $\Delta$ to the left and right of the junction by $e^{\mp i\phi/2}$, respectively. Despite the simplicity of this description, it contains the relevant physics of a short SNS junction. As it has been shown for standard junctions \cite{Cuevas:PRB96}, such physics essentially depend on the contact normal transmission as well as the voltage drop across it \footnote{Note that, for the sake of simplicity, we do not include the possibility of junctions containing resonant levels or quantum dots. A study of such junctions, including Coulomb blockade effects, is beyond the scope of this paper but might be useful in order to analyse the possibility of Majorana physics arising in experiments with short SNS junctions containing quantum dot nanowires, such as the ones reported in Ref. \cite{Deng:NL12}}. Thus, we expect that a more detailed modeling, including e.g. a spatially-dependent voltage drop, would only modify the effective transmission $T_N(\nu)$ which defines the different regimes we shall explain in the following.

\subsection{Andreev bound states}\label{ABS-SNS}
In such short SNS junction, an ABS should form for each of the two p-wave sectors described in section \ref{Rashba NW} for $B<B_c$, while only one, associated to $\Delta_+$, should remain for $B>B_c$. To support this picture, we present calculations of the local density of states (LDOS) at the junction in the transparent limit ($T_N=1$). This LDOS is peaked at the energy $\epsilon_\pm$ of the ABS, which is a function of the phase difference $\phi$ across the junction. For $B=0$ (Fig. \ref{fig:ABS}a) the two ABSs are degenerate and confined within the gap $\Delta$ \footnote{Note that even this non-topological case is anomalous as the ABS energies do not reach zero at $\phi=\pi$, unlike predicted by the standard theory for a transparent channel $T_N=1$ within the Andreev approximation $\mu\gg \Delta$ \cite{Beenakker:92}. We have checked that the energy minimum $\delta_\pi$ does indeed vanish as $\mu/\Delta$ grows, see Fig. \ref{fig:deltapi} in \ref{AppendixC}}. As the Zeeman field increases, Fig. \ref{fig:ABS}b, the two ABS split and the system develops the two distinct gaps $\Delta_+$ and $\Delta_-$ described in Section \ref{Rashba NW}. Note that both ABSs are truly bound at energies below the lowest gap $\Delta_-$, but only  \emph{quasibound}  in the energy window $\Delta_-<\epsilon<\Delta_+$. This is readily apparent in the plot as a broadening of the ABS resonances (see, e.g.  Fig. \ref{fig:ABS}c). As $B$ approaches the critical field $B_c$, $\Delta_-$ gets reduced, and becomes exactly zero at $B=B_c$. Note that at this point the upper ABS has reached zero energy at  $\phi=\pi$ and is quasibound for all energies, Fig. \ref{fig:ABS}d. Upon entering the topological phase ($B\geq B_c$), $\Delta_-$ reopens but one of the ABSs of the problem \emph{has disappeared} (Fig. \ref{fig:ABS}e). The surviving ABS associated to $\Delta_+$ arises due to the hybridization of the two emerging MBSs across the junction. Global fermion-parity conservation protects the $\phi=\pi$ level crossing. Due to the residual $\Delta^{+-}_s$ coupling between the two sectors, the $\Delta_+$ Andreev state is once more quasibound in the energy window $\Delta_-<\epsilon<\Delta_+$. At high enough magnetic fields, $\Delta_+$ is the smallest gap of the problem and hence the Majorana ABS is truly bound, Fig. \ref{fig:ABS}f. In long junctions, more ABSs can be confined in the junction \cite{Prada:PRB12,Cayao:UP,Chevallier:PRB12}. These extra ABSs coexist with the ones described here and may, for example, anti-cross with the Majorana-like $\Delta_+$ Andreev level, affecting its character near zero energy \cite{Prada:PRB12}.
\subsection{ac Josephson effect and MAR currents}\label{ac Josephson}
\begin{figure}[t] 
   \centering
   \includegraphics[width=\linewidth,clip]{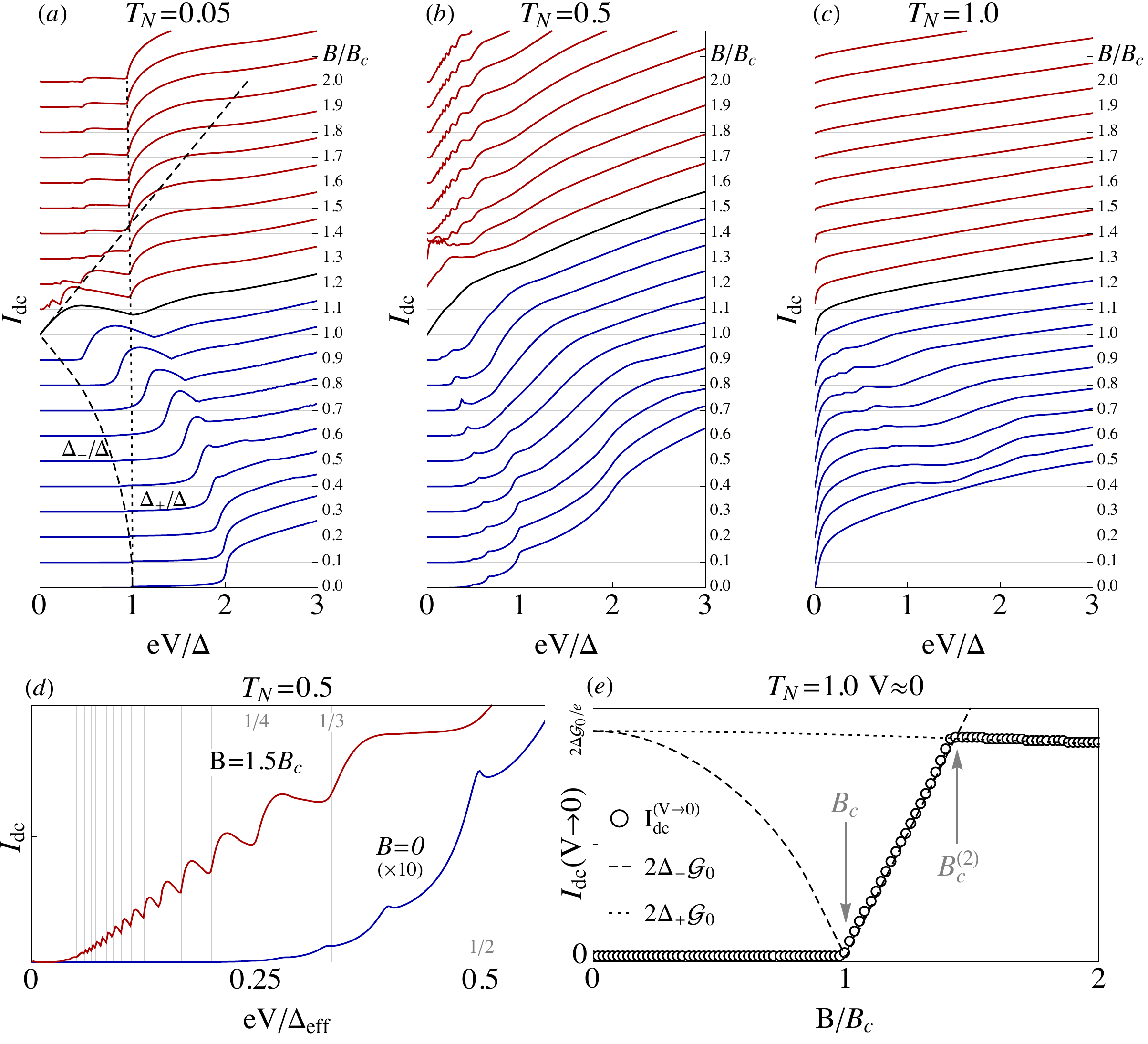} 
     \caption{(Color online) Time-averaged Josephson current $I_{dc}$ as a function of bias $V$ for increasing Zeeman field $B$. Curves are offset by a constant $2\Delta \mathcal{G}_0/e$, with $\mathcal{G}_0=e^2/h$. Blue and red curves correspond to the non-topological ($B<B_c$) and topological ($B>B_c$) phases respectively. Panels (a) to (c) show the cases of tunnelling, intermediate and full transparency.  Panel (d) is a blowup of the low bias MAR subharmonics at intermediate transparency. Panel (e) shows the asymptotic $I_{dc}(V\to 0)$ at full transparency (circles), along with the dependence of the quantities $2\Delta_- \mathcal{G}_0$ and $2\Delta_+\mathcal{G}_0$ with $B$ across the topological transition  [dashed/dotted lines, evolution also shown in panel (a)].
}
   \label{fig:I0}
\end{figure}

Under a constant voltage bias $V$, the pairings $\Delta$ to the left and right of the junction acquire an opposite and time-dependent phase difference, $\phi(t)=2 eVt/\hbar$. This induces Landau-Zener transitions between the ABSs and into the continuum, thereby developing a time dependent Josephson current with both $I_{dc}$ and $I_{ac}$ components. Such is the point of view in e.g. Refs. \cite{Averin:PRL95,San-Jose:PRL12a}. 
Alternatively, $\phi(t)$ can be gauged away into the hopping across the junction, $v_0(t)=\nu e^{-i\frac{eV}{\hbar} t\tau_z}\sum_{\sigma\sigma'} c^+_{r\sigma}v_{\sigma\sigma'}c_{l\sigma'} + \mathrm{h.c.}$, where $\tau_z$ is the $z$-Pauli matrix in Nambu space. 
By employing Keldysh-Floquet theory \cite{Cuevas:PRB96, Sun:PRB02}, we obtain the stationary-state time-dependent ac Josephson current $I(t)=\sum_n e^{in\frac{eV}{\hbar} t}I_n$ (note that only even harmonics survive, see \ref{AppendixA} for full details).  Here, we concentrate on the dc-current $I_{dc}=I_0$. The results for $I_{dc}(V)$ at small, intermediate and full transparency are summarised in Fig. \ref{fig:I0}(a-c) for increasing values of $B$ spanning the topological transition.

\subsubsection{Tunneling regime.} For non-topological tunnel junctions, dc-transport vanishes below an abrupt threshold voltage $V_t=2\Delta_\mathrm{eff}/e=2\Delta_-/e$ (Fig. \ref{fig:I0}(a), blue curves). This well known result follows from the fact that there are no quasiparticle excitations in the decoupled wires for energy $\epsilon\in(-\Delta_\mathrm{eff},\Delta_\mathrm{eff})$ if $B<B_c$. Indeed, to second order in perturbation theory in $\nu$, the MAR current takes the form of a convolution between $A_0(\omega)$ and $A_0(\omega\pm eV/\hbar)$, where $A_0$ is the decoupled ($\nu=0$) spectral density at each side of the junction (\ref{AppendixB}). [The trace of $A_0(\epsilon)$, proportional to the local density of states (LDOS), is shown in Fig. \ref{fig:Ic}(a)]. 
Hence, as $B$ increases, the tunnelling current threshold follows the closing of the gap in the LDOS, until $V_t$ vanishes and $I_{dc}$ becomes linear in small $V$ at $B_c$ (black curve). As $B>B_c$, the gap reopens, but the threshold is now halved to $V_t=\Delta_\mathrm{eff}/e$ (red curves) \footnote{The small step visible at $eV=\Delta_\mathrm{eff}/2$ is the second-order MAR, whose relative height vanishes as $T_N\to 0$}. The change, easily detectable as a halving of the slope of the threshold $dV_t/dB$ across $B_c$, is due to the emergence of an intra-gap zero-energy MBS in the topological phase [see zero energy peak in Fig. \ref{fig:Ic}(a)], which opens a tunnelling transport channel from or into the new zero energy state. Moreover, when $B=B_c^{(2)}$, $\Delta_-$ surpasses $\Delta_+$, and $\Delta_\mathrm{eff}$ saturates at $\Delta_+$. This is directly visible in $V_t(B)$ as a kink at $B_c^{(2)}$ [see dashed and dotted lines in Fig. \ref{fig:I0}(a)]  \footnote{Similar considerations may apply to recent experiments with lead nanoconstrictions formed in an STM tip, see Ref. \cite{Rodrigo:PRL12}.}. 

\subsubsection{Intermediate transparency regime.} As transparency increases, subharmonic MAR steps develop at voltages $V_t/n=2\Delta_\mathrm{eff}/en$ ($n=2,3,4,\dots$), see Fig. \ref{fig:I0}(b).
The specific profile of each step with $V$ still contains information on the LDOS of the junction at energies around $\Delta_\mathrm{eff}$. At $B=0$, the power-law LDOS for $|\epsilon|>\Delta$ results in a staircase-like curve $I_{dc}(V)$ 
[blue line in Fig. \ref{fig:I0}(d)]. This shape is roughly preserved up to $B=B_c$. For $B>B_c$ the MAR profile changes qualitatively, however. The subharmonic threshold voltages $V_t/n$ are halved (since $V_t$ is halved), and the MAR current profile becomes oscillatory instead of step-like. A blowup of the oscillations is presented in Fig. \ref{fig:I0}(d) (red curve), together with guidelines for the corresponding $V_t/n$ in gray. 


The emergence of oscillatory MAR steps, which here is connected to the formation of  zero energy peaks in the LDOS owing to the localized MBSs, is well known in Josephson junctions containing a resonant level \cite{Johansson:PRB99,Yeyati:PRL03,Jonckheere:PRB09}. Note, however, that the oscillations in a topologically trivial system,  such as for instance a quantum dot between two superconductors, arise at odd fractions of $2\Delta_\mathrm{eff}$, i.e. at voltages $2\Delta_\mathrm{eff}/(2n-1)e$, instead of the $\Delta_\mathrm{eff}/en$ of the Majorana case. Interestingly, this difference is ultimately due to the fact that a resonant level spatially localised within the junction cannot carry current directly into the reservoirs, while a zero energy MBS (essentially half a non local fermion) can. This same situation arises in $d$-wave Josephson junctions, which also exhibit oscillatory $\Delta_\mathrm{eff}/en$ MAR subharmonics owing to the presence of mid gap states \cite{Cuevas:PRB01}.

\subsubsection{Transparent limit.} In the limit $T_N\to 1$, ABS energies $\epsilon_\pm(\phi)$ [Fig. \ref{fig:ABS}(c)] touch the continuum at $\phi=0$. This has an important consequence. From the Landau-Zener point of view of the ac Josephson effect \cite{Averin:PRL95}, the time dependence of $\phi(t)=2eV t/\hbar$ for an arbitrarily small $V$ will induce the escape of any quasiparticle occupying an ABS into the continuum after a single $\phi(t)$ cycle. A given ABS becomes occupied with high probability in each cycle around $\phi=\pi$ if the rate $\hbar\,d\phi(t)/dt=2eV$ exceeds its energy minimum $\epsilon(\pi)\equiv \delta_\pi$. (Recall this energy is finite, since the Andreev approximation does not apply, see \ref{AppendixC}.) 
One quasiparticle is then injected into the continuum per cycle, and a finite $I_{dc}(V\gtrsim \delta_\pi/e)$ arises. Below such voltage, however, the ABS remains empty, so that if $\delta_\pi$ is finite, as is the case of a realistic non-topological junction [see Fig. \ref{fig:ABS}(a-c)], one obtains $I_{dc}(V\to 0)=0$ (valid for any transparency at $B<B_c$). This is in contrast to the conventional $B=0$, $T_N=1$ result $I(V\to 0)=4\Delta\mathcal{G}_0/e$, predicted within the Andreev approximation ($\mathcal{G}_0=e^2/h$).


After the topological transition, this picture changes dramatically. The two MBSs at each side of the junction hybridise for a given $\phi$ into a \emph{single} ABS. This seemingly innocent change has a notable consequence. Since fermion parity in the superconducting wires is globally preserved, an anticrossing at $\phi=\pi$, which would represent a mixing of a state with one and zero fermions in the lone ABS, is forbidden. Parity conservation therefore imposes $\delta_\pi=0$ in the presence of MBSs, irrespective of $T_N$ or $\mu/\Delta$ \footnote{Note that residual splitting may survive in the topological phase for finite length nanowires, for which a finite (albeit exponentially small) coupling between four MBSs exist.}. This is a true topologically protected property of the junction, and gives rise to a finite $I_{dc}(V\to 0)=2\Delta_\mathrm{eff} \mathcal{G}_0/e$, i.e.  half the value expected for the  non-topological junction in the Andreev approximation. This abrupt change is shown in Fig. \ref{fig:I0}(c,e).
 The $I_{dc}(V\to 0)$ MAR current in transparent junctions, therefore, directly probes the emergence of parity protection.

\begin{figure}[t] 
   \centering
   \includegraphics[width=0.75\linewidth,clip]{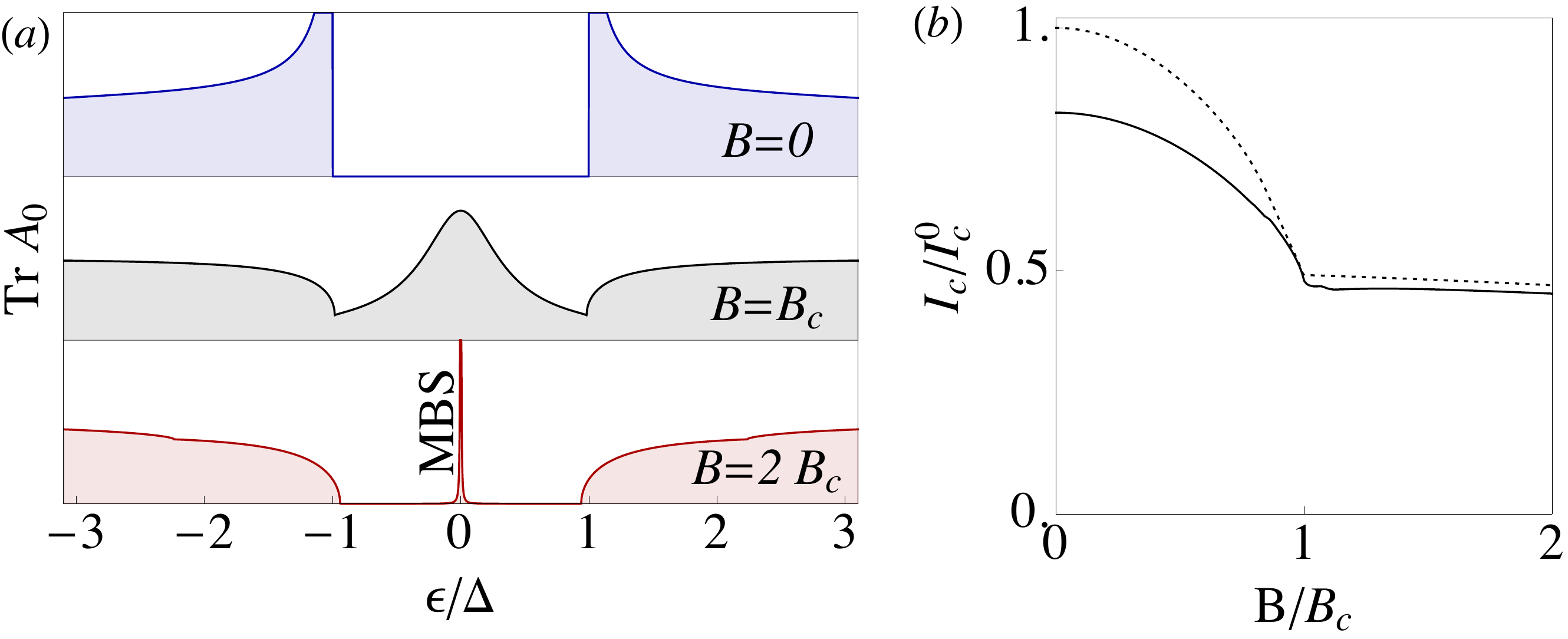} 
     \caption{(Color online) (a) Local density of states at the end of a single nanowire in the non-topological (top), critical (middle) and topological phase (bottom). A zero-energy Majorana peak appears in the latter case. (b) The critical current $I_c(B)$ for $T_N=1$ across the topological transition in units of $I_c^0=e\Delta/\hbar$. 
     The dotted line corresponds to $\frac{1}{2}(\Delta_+ + \Delta_-)/\Delta$ for $B<B_c$, and $\frac{1}{2}\Delta_+/\Delta$ for $B>B_c$. 
}
   \label{fig:Ic}
\end{figure}

\section{Critical current}\label{Critical current} In the transparent limit, a supercurrent peak  \cite{Doh:S05,Nilsson:NL12, Chauvin:PRL07} may hinder the experimental identification of the $I_{dc}(V\to 0)$ limit, but itself holds valuable information about the transition.  
The critical current $I_c$ may be computed in general by maximizing the $V=0$ (time-independent) current $I(\phi)$ respect to $\phi$ (including the contribution from the continuum). For a short transparent junction at $B=0$, $I_c$ is maximum, and equal to $I_c^0\equiv e\Delta/\hbar$ in the Andreev approximation.
 Fig. \ref{fig:Ic}(b) shows $I_c$ for increasing values of $B$. Naively, one may expect that a junction without a superconducting gap should not carry a finite supercurrent, but this is not the case here. At $B=B_c$, $I_c$ is finite \footnote{Note also that in junctions with trivial superconductors, $I_c\rightarrow 0$ as the nanowire becomes helical at $B=\mu<B_c$ \cite{Cheng:PRB12}. Our result is therefore another nontrivial consequence of topology in the junction.}, while $\Delta_\mathrm{eff}=0$ [the junction LDOS at criticality is also gapless, see Fig. \ref{fig:Ic}(a)]. This gapless supercurrent comes from the $\epsilon_+(\phi)$ \emph{quasi-bound} Andreev state in the continuum, which contributes almost as if it were a subgap ABS. It is thus a reasonable approximation to write $I_c$ as the sum of the critical current from each ABS. For $B<B_c$, $I_c\approx \frac{1}{2}I_c^0(\Delta_++\Delta_-)/\Delta$. The $\Delta_-$ contribution, however, should not be included for $B>B_c$, leading to a discontinuity in $\partial I_c/\partial B$. This simple model gives a qualitative fit [dotted line in Fig. \ref{fig:Ic}(b)] to the exact numerics (solid line), with deviations coming from corrections to the Andreev approximation, and contributions above $\Delta_+$. Additional deviations in experiments, coming e.g. from the finite impedance of the electromagnetic environment, are not expected to alter the discontinuity in $\partial I_c/\partial B$, which remains a signature of the  topological transition. 
 
%
 
\section{Conclusions}
In conclusion, we have shown that the dc-current in voltage biased Josephson junctions is a flexible experimental probe into the various aspects of the topological superconducting transition in semiconducting nanowires. Tuning the junction transparency one may obtain evidence of MBS formation as conclusive as a fractional Josephson effect, without requiring control of the junction phase. Moreover, we have found that the critical current in the wire does not vanish at the transition due to above-gap contributions, although its derivative with $B$ exhibits a discontinuity as a result of the disappearance of one ABS. This behavior of $I_c$ provides a direct evidence of the topological transition. MAR spectroscopy and critical current measurements in nanowires similar to the ones studied here have already been reported \cite{Doh:S05,Nilsson:NL12}. 

Although we have focused here on the simplest case (single-band, short junction limit) we expect the main features of the topological transition to remain robust under more general conditions. Preliminary results in the quasi-one dimensional multiband case show that $I_c$ is a non-monotonic function for increasing magnetic fields. For weak interband SO mixing \cite{Lim:PRB12}, the behavior discussed in Fig. \ref{fig:Ic}(b) can be generalized and $I_c$  presents a series of minima at different fields corresponding to the topological transition of each subband.

Importantly, the alternative physical scenarios, such as, e. g., disorder \cite{Pientka:PRL12,Bagrets:PRL12,Liu:PRL12,Sau:13} or Andreev bound states  \cite{Lee:13}, that produce ZBAs in NS junctions (and thus mimic Majorana physics), cannot give the distinct features associated to global parity that were discussed here for SNS junctions. We therefore believe that experiments along the lines discussed in this paper could provide the first unambiguous report of a topological transition in nanowires, and the emergence of Majorana bound states.

\ack We acknowledge the support of the European Research Council, the Spanish Research Council CSIC through the JAE-Predoc Program (J. C.) and the Spanish Ministry of Economy and Innovation through Grants No. FIS2012-33521, FIS2011-23713, FIS2010-21883, FIS2009-08744 and the Ramón y Cajal Program (E. P). 

\appendix
\section{Floquet-Keldysh formalism}\label{AppendixA}
Consider a mesoscopic system composed of two semi-infinite leads (labeled $L$ and $R$), each in thermal equilibrium at the same temperature $T$ and with the same chemical potential $\mu=0$. Each lead has a finite s-wave superconducting pairing $\Delta_\alpha$, where $\alpha=L,R$. A central system ($\alpha=S$), which may or may not be superconducting, is coupled to both leads through operator $v$. In its Nambu form, the Hamiltonian of the system reads
\[
\hat H=\frac{1}{2}\sum_{ij}\left(\begin{array}{c|c}
c_j&c^+_j
\end{array}\right)
H_{ij}\left(\begin{array}{c}
c_j\\\hline c^+_j
\end{array}\right),
\]
where the Nambu Hamiltonian matrix takes the general form
\[
H=\left(\begin{array}{ccc|ccc}h_L & v^+ & 0 & \Delta_L & 0 & 0 \\ v& h_S & v^+ & 0 & \Delta_S & 0 \\ 0 & v & h_R & 0 & 0 & \Delta _R\\\hline \Delta_L^+ & 0 & 0 & -h_L^* & (-v^+)^* & 0 \\ 0 & \Delta_S^+ & 0 & -v^* & -h_S^* & (-v^+)^* \\ 0 & 0 & \Delta_R^+ & 0 & -v^* & -h_R^*\end{array}\right).
\]
Here $h_\alpha$ is the normal Hamiltonian for each section of the system. The blocks delimited by lines denote the Nambu particle, hole and pairing sectors.

If we apply a left-right voltage bias $V$ through the junction, the Bardeen-Cooper-Schrieffer (BCS) pairing of the leads will become time dependent, $\Delta_{L/R} \to e^{\pm i Vt}\Delta_{L/R}$, while $h_{L/R}\to h_{L/R}\pm V/2$ (we take $e=\hbar=1$). Both these changes can be gauged away from the leads and into the system by properly redefining $c^+_i\to c^+_i(t)= e^{\pm iVt/2} c^+_i$. This transformation is done also inside the system $S$, thereby effectively dividing it into two, the portion with an $e^{iVt/2}$ phase (denoted $S_L$), and the portion with the opposite phase (denoted $S_R$). This restores $H$ to its unbiased form, save for a new time dependence in $h_S\to h_S(Vt)$, which is constrained to the coupling between the $S_L$ and $S_R$,
\[
h_S(Vt)=\left(\begin{array}{cc}h_{S_L} & e^{-iVt}v^+_0\\ e^{iVt}v_0 & h_{S_R} \end{array}\right).
\]  
It is important to note that $H(t)$ is periodic, with angular frequency $\omega_0=V$. In the steady state limit (at long times $t$ after switching on the potential $V$) all response functions and observables will exhibit the same time periodicity (all transient effects are assumed to be completely damped away). In particular, the steady state current $I(t)=I(t+2\pi/\omega_0)$, so that
\[
I(t)=\sum_n e^{in\omega_0 t}I_n\,,
\]
for some harmonic amplitudes $I_n$, in general complex, that satisfy $I_n=I_{-n}^*$ since $I(t)$ is real.
This current can be computed using the Keldysh Green's function formalism. \cite{Haug:07} The standard expression for $I(t)$ is computed starting from the definition of $I(t)=\partial_t N_L$, where $N_L$ is the total number of fermions in the left lead. By using Heisenberg equation and the Keldysh-Dyson equation, one arrives at
\[
I(t)=\mathrm{Re}[J(t)],
\]
where
\[
J(t)=\frac{2e}{\hbar} \int dt'~\mathrm{Tr}\left\{\left[
G^r(t,t')\Sigma_L^<(t',t) + G^<(t,t')\Sigma_L^a(t',t)\right]\tau_z\right\}.
\]
The z-Pauli matrix $\tau_z$ above acts on the Nambu particle-hole sector,
\[
\tau_z=\left(\begin{array}{c|c}\mathbbm{1} &0\\\hline 0 & -\mathbbm{1} \end{array}\right).
\] 
The self energy from the left lead is defined as $\Sigma_L^{a,<}(t',t)=v\,g_L^{a,<}(t',t) v^+$, where $g_L(t',t)=g_L(t'-t)$ stands for the left lead's propagator, when decoupled from the system (this propagator depends only on the time difference since the decoupled lead is time independent in this gauge).
We define the Fourier transform of $g$ as
\[
g(\omega)=\int_{-\infty}^\infty dt e^{i\omega t} g(t).
\]
The retarded propagator in Fourier space is 
\[
g_L^r(\omega)=\frac{1}{\omega-h_L+i\eta}\,,
\]
while the advanced $g^a_L(\omega)=\left[g^r_L(\omega)\right]^+$. One can compute $g_L^<(\omega)=i f(\omega) A_L(\omega)$, where  $f(\omega)=1/(e^{\omega/k_BT}+1)$ is the Fermi distribution in the leads, and $A_L(\omega)=i(g_L^r(\omega)-g_L^a(\omega))$ is the Nambu spectral function. The $g^r_{L/R}$ [and in particular $A_{L/R}(\omega)$] is assumed known, or at least easily obtainable from $h_{L/R}$ and $v$. 
Finally, the Green functions $G^r(t',t)$ and $G^<(t',t)$ correspond to the propagator for the full system, including the coupling to the leads. (Note that, in practice, since $G$ is inside a trace in $J(t)$, only matrix elements of $G$ \emph{inside} the $S$ portion of the full system are needed).
The retarded $G^r$ satisfies the equation of motion
\[
\left[i\partial_{t'}-H(t')\right]G(t',t)=\delta(t'-t),
\]
while $G^<$ (when projected onto the finite-dimensional system $S$) satisfies the Keldysh relation
\begin{eqnarray*}
G^<(t',t)&=&\int dt_1dt_2 G^r(t',t_1)\\
&&\times\left[\Sigma_L^<(t_1-t_2)+\Sigma_R^<(t_1-t_2)\right]G^a(t_2,t).
\end{eqnarray*}
Since $h_S$ in $H$ is time dependent, $G$ propagators depend on two times; unlike $\Sigma_{L/R}$ or $g_{L/R}$ they are not Fourier diagonal. Instead, we can exploit the steady-state condition, which reads
\[
G(t',t)=G(t'+\frac{2\pi}{\omega_0},t+\frac{2\pi}{\omega_0}),
\]
to expand the system's $G$ as a Fourier transform in $t'-t$ and a Fourier \emph{series} in $t$. We define
\[
G(t',t)=\sum_n e^{-in\omega_0 t}\int_{-\infty}^\infty \frac{d\epsilon}{2\pi}e^{-i \epsilon(t'-t)}G_{n}(\epsilon).
\]
The natural question is how the equation of motion is expressed in terms of the harmonics $G_n(\epsilon)$. It takes the most convenient form if we redefine $G_n(\epsilon)$ (where $\epsilon$ is unbounded) in terms of the quasienergy $\tilde{\epsilon}\in [0,\hbar\omega_0]$, i.e. $\epsilon=\tilde\epsilon+m\omega_0$
\[
G_{mn}(\tilde\epsilon)=G_{m-n}(\tilde\epsilon+m\omega_0).
\]
This has the advantage that the equation of motion translates to a matrix equation analogous to that of a static system in Fourier space
\[
\sum_m (\tilde \epsilon+n'\omega_0-H_{n'm})G^r_{mn}(\tilde\epsilon)=\delta_{n'n},
\]
where 
\[
H_{n'n}=\int dt e^{i(n'-n)t} H(t).
\] 
This is known as the Floquet description of the steady state dynamics in terms of sidebands, which appear formally as a new quantum number $n$. Time dependent portions of $H(t)$ act as a coupling between different sidebands. The effective Hamiltonian for the $n$-th sideband is the static portion of $H(t)$, shifted by $-n\omega_0$. One therefore sometimes defines the Floquet ``Hamiltonian'' of the system as 
\[
{\bm{h_S}}_{nm}={h_S}_{nm}-n\omega_0\delta_{nm},
\]
where, as before, ${h_S}_{n'n}=\int dt e^{i(n'-n)t} {h_S}(t)$. Likewise, one may define the Floquet self-energies as 
\[
{\bm{\Sigma_L}}_{nm}(\tilde\epsilon)=\delta_{nm}\Sigma_{L/R}(\tilde\epsilon+n\omega_0),
\]
(since the leads are static, $\bm{\Sigma}$ is sideband-diagonal).

The Floquet equation of motion for $G^r_{nm}(\tilde\epsilon)$ can be solved like in the case of a static system. Within the $S$ portion of the system, we have
\[
\bm{G}^r(\tilde\epsilon)=\left[\tilde\epsilon-\bm{h_S}-\bm{\Sigma_L}^r(\tilde \epsilon)-\bm{\Sigma_R}^r(\tilde \epsilon)\right]^{-1}.
\]
Boldface denotes the sideband structure implicit in all the above matrices.
Similarly, the Keldysh relation takes the simple form
\[
\bm{G^<}(\tilde \epsilon)=\bm{G^r}(\tilde \epsilon)\left[\bm{\Sigma_L}^<(\tilde \epsilon)+\bm{\Sigma_R}^<(\tilde \epsilon)\right]\bm{G^a}(\tilde \epsilon).
\]
Finally, the time averaged current $I_{dc}\equiv I_0$ takes the form
\begin{equation}
\label{KeldyshIdc}
I_{dc}=\frac{2e}{h} \int_0^{\hbar\omega_0} d\tilde\epsilon~\mathrm{Re}\mathrm{Tr}\left\{\left[\bm{G^r}(\tilde\epsilon)\bm{\Sigma_L^<}(\tilde\epsilon)+\bm{G^<}(\tilde\epsilon)\bm{\Sigma_L^a}(\tilde\epsilon)\right]\tau_z\right\},
\end{equation}
where the trace includes the sideband index. 
In a practical computation, the number of sidebands that must be employed is finite, and depends on the applied voltage bias $V$ (the typical number scales as $n_\mathrm{max}\sim v_0/V$). We employ an adaptive scheme that increases the number of sidebands recursively until convergence for each value of $V$.

\section{Tunneling limit}\label{AppendixB}
\begin{figure}[t] 
   \centering
   \includegraphics[width=0.3\linewidth,clip]{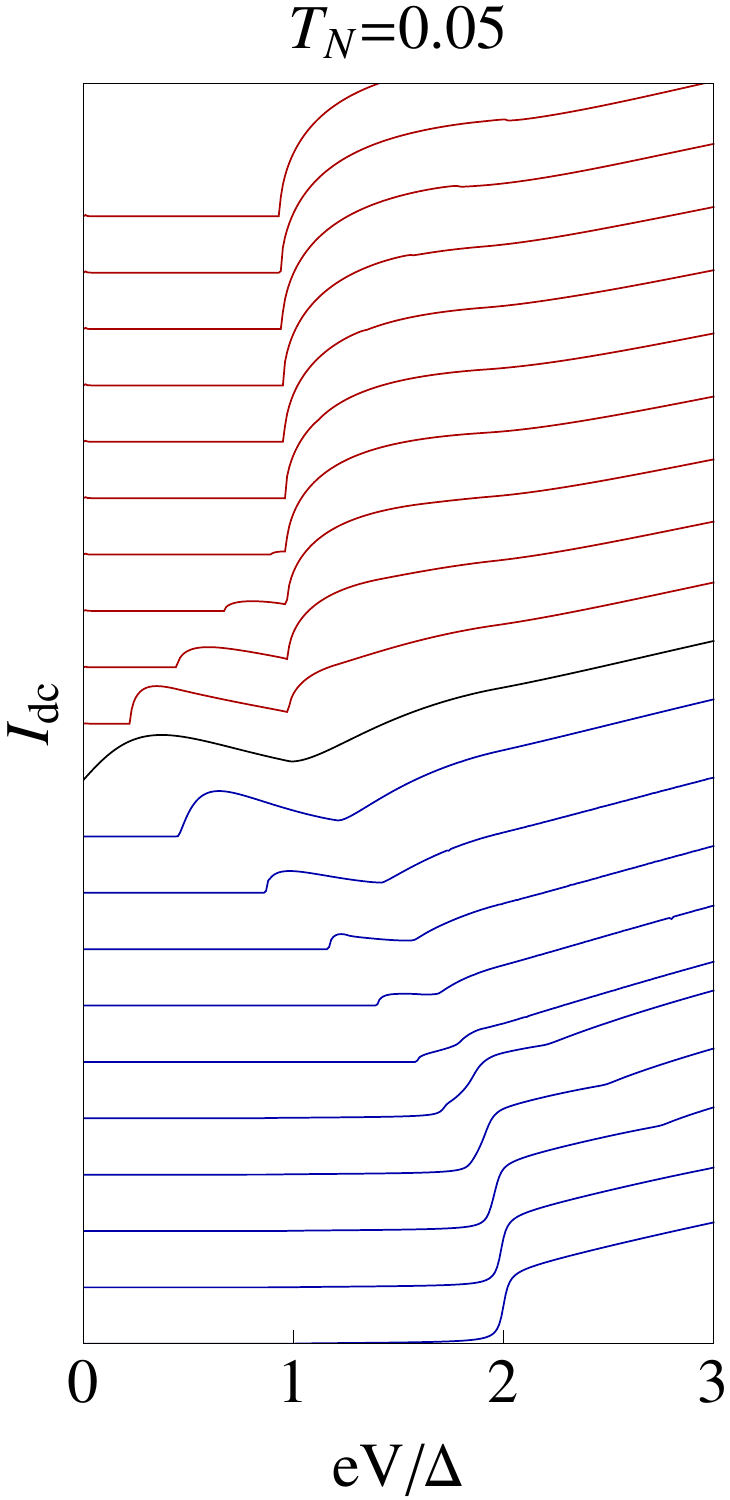} 
     \caption{(Color online) Time-averaged current $I_{dc}$ as a function of bias $V$ for increasing Zeeman field $B$ using the tunneling approximation of Eq. \ref{Itunnel}. Curves are offset by a constant $2\Delta \mathcal{G}_0/e$, with $\mathcal{G}_0=e^2/h$. Blue and red curves correspond to the non-topological ($B<B_c$) and topological ($B>B_c$) phases respectively. 
}
   \label{fig:Itun}
\end{figure}

It is possible to solve the $I_{dc}$ current explicitly in the tunnelling limit. To leading (second) order in the left-right coupling $v_0$, Eq. (\ref{KeldyshIdc}) reduces, after some algebra, to
\begin{eqnarray}
I_{dc}&\approx&\frac{e}{\pi} \mathrm{Re}\int d\omega \left[f(\omega-\omega_0)-f(\omega)\right]
\mathrm{Tr}\left\{A_{11}^L(\omega) v_0^+ A_{11}^R(\omega-\omega_0)v_0\right\}
\label{Itunnel}
\end{eqnarray}
where $A_{11}^\alpha$ is the particle-particle Nambu $2\times2$ matrix block of the spectral function of the $\alpha=L,R$ decoupled wire,
\[
A^{\alpha}(\omega)
=\left(\begin{array}{cc}
A_{11}^\alpha(\omega) & A_{12}^\alpha(\omega)\\
\left[A_{12}^\alpha(\omega)\right]^+ & -\left[A_{11}^{\alpha}(-\omega)\right]^*
\end{array}\right),
\]
and the trace is taken over spin space. The trace of $A^{\alpha}(\omega)$ is proportional to the local density of states. Fig. \ref{fig:Itun} shows results for the tunnel current using Eq. \ref{Itunnel} for increasing Zeeman fields. Overall, the agreement with the full numerics in Fig. \ref{fig:I0}(a) is very good and, importantly, all the relevant features such as, e. g.,  the closing of the gap, are captured by this tunneling approximation.

\section{Andreev approximation}\label{AppendixC}

It is conventional, in the study of hybrid superconducting-normal junctions, to assume the limit in which the Fermi energy $\mu$ of the metal under consideration is much greater than the superconducting gap, and any other energy $E$ involved in the problem, $\mu\gg\Delta, E$. This is known as the Andreev approximation. In essence, it allows one to regard the normal system as featureless, with constant Fermi velocity and density of states. In this case, a number of simplifications can be carried out in the computation of equilibrium transport properties. One important consequence of the approximation in the context of our work is that, in a short SNS junction with phase difference $\phi$, and symmetric under time-reversal symmetry ($B=0$ in our case), two degenerate Andreev states will appear of energy $\epsilon(\phi)=|\Delta|\sqrt{1-T_N^2\sin^2(\phi/2)}$ \cite{Beenakker:92}. 
This immediately implies that at perfect normal transparency $T_N=1$, the ABSs will reach zero energy at $\phi=\pi$. In other words, in the Andreev approximation the ABS energy minimum in the non-topological phase will be $\epsilon(\pi)=\delta_\pi=0$.  

While in the topological phase, a zero $\delta_\pi$ is a robust property, protected by parity conservation, a $\delta_\pi=0$ in the non-topological phase is accidental, a direct consequence of the Andreev approximation, and is not protected by any symmetry. In fact, any deviation from the Andreev approximation will induce a finite splitting $\delta_\pi$. In semiconducting nanowires such as the ones considered in this work, this correction is very relevant. Indeed, for the nanowire to undergo a superconducting topological transition at a reasonable Zeeman field $B$, $\mu/\Delta$ must remain relatively small (the wire must be close to depletion, and far from the Andreev approximation), otherwise the critical field $B_c=\sqrt{\mu^2+\Delta^2}$ would be physically unreachable. The splitting $\delta_\pi$, therefore, remains a relevant quantity in the formation and detection of Majorana bound states.

\begin{figure}[t] 
   \centering
   \includegraphics[width=0.5\linewidth,clip]{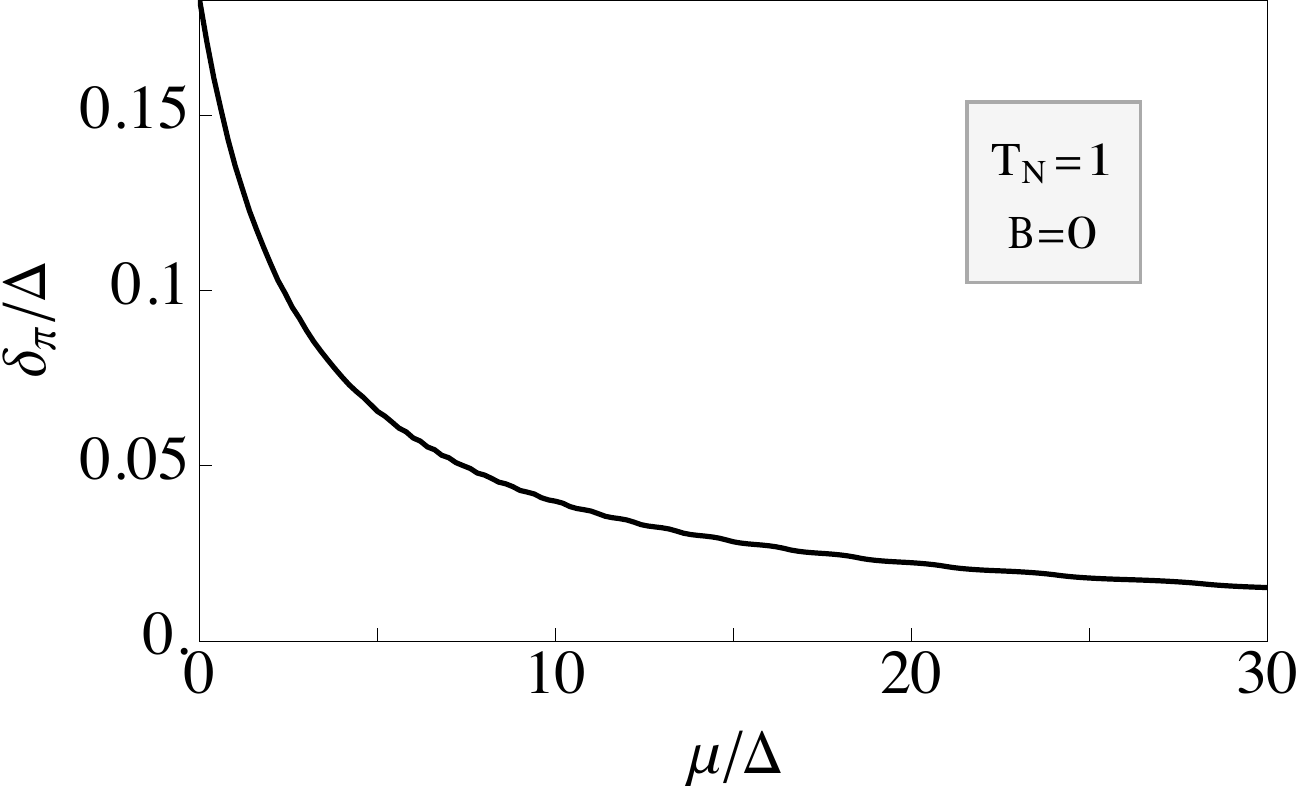} 
     \caption{Minimum Andreev state energy $\delta_\pi$ in a short SNS junction, at $B=0$ and $T_N=1$. As $\mu/\Delta$ grows, $\delta_\pi$ decreases to zero, in agreement with predictions within the Andreev approximation.
}
   \label{fig:deltapi}
\end{figure}

The value of $\delta_\pi$ in our system may be computed numerically. The most efficient way is to consider a short SNS junction with finite length superconductors, $T_N=1$, $B=0$ and a phase difference $\phi=\pi$. Since this system is closed, an exact diagonalization of the tight-binding Nambu Hamiltonian yields a minimum eigenvalue that is exactly $\delta_\pi$ if the S leads are long enough (longer than the coherence length). We find that this quantity is finite in the case of wires close to depletion, $\mu\gtrsim\Delta$, and that it vanishes as one approaches the Andreev approximation regime $\mu\gg\Delta$, see Fig. \ref{fig:deltapi}. More specifically, we have found that $\delta_\pi$ scales as $\delta_\pi=c_1 \Delta^2/(\mu+\Delta c_2)$ for some  $c_{1,2}>0$, within very good precision.

\section*{References}

\bibliographystyle{unsrt}

\bibliography{biblio}

\begin{thebibliography}{10}

\bibitem{Lutchyn:PRL10}
Roman~M. Lutchyn, Jay~D. Sau, and S.~Das~Sarma.
\newblock Majorana fermions and a topological phase transition in
  semiconductor-superconductor heterostructures.
\newblock {\em Phys. Rev. Lett.}, 105(7):077001, Aug 2010.

\bibitem{Oreg:PRL10}
Yuval Oreg, Gil Refael, and Felix von Oppen.
\newblock Helical liquids and majorana bound states in quantum wires.
\newblock {\em Phys. Rev. Lett.}, 105:177002, Oct 2010.

\bibitem{Delft-exp}
V.~Mourik, K.~Zuo, S.~M. Frolov, S.~R. Plissard, E.~P. A.~M. Bakkers, and L.~P.
  Kouwenhoven.
\newblock Signatures of majorana fermions in hybrid
  superconductor-semiconductor nanowire devices.
\newblock {\em Science}, 336:1003, 2012.

\bibitem{Deng:NL12}
M.~T. Deng, C.~L. Yu, G.~Y. Huang, M.~Larsson, P.~Caroff, and H.~Q. Xu.
\newblock Anomalous zero-bias conductance peak in a nb--insb nanowire--nb
  hybrid device.
\newblock {\em Nano Lett.}, 12(12):6414--6419, 2012.

\bibitem{Das:NP12}
Anindya Das, Yuval Ronen, Yonatan Most, Yuval Oreg, Moty Heiblum, and Hadas
  Shtrikman.
\newblock Zero-bias peaks and splitting in an al-inas nanowire topological
  superconductor as a signature of majorana fermions.
\newblock {\em Nat Phys}, 8(12):887--895, 12 2012.

\bibitem{Sengupta:PRB01}
K.~Sengupta, Igor Zutik, Hyok-Jon Kwon, Victor~M. Yakovenko, and S.~Das~Sarma.
\newblock Midgap edge states and pairing symmetry of quasi-one-dimensional
  organic superconductors.
\newblock {\em Phys. Rev. B}, 63:144531, Mar 2001.

\bibitem{Bolech:PRL07}
C.~J. Bolech and Eugene Demler.
\newblock Observing majorana bound states in $p$-wave superconductors using
  noise measurements in tunneling experiments.
\newblock {\em Phys. Rev. Lett.}, 98:237002, Jun 2007.

\bibitem{Law:PRL09}
K.~T. Law, Patrick~A. Lee, and T.~K. Ng.
\newblock Majorana fermion induced resonant andreev reflection.
\newblock {\em Phys. Rev. Lett.}, 103:237001, Dec 2009.

\bibitem{Prada:PRB12}
Elsa Prada, Pablo San-Jose, and Ram\'on Aguado.
\newblock Transport spectroscopy of $ns$ nanowire junctions with majorana
  fermions.
\newblock {\em Phys. Rev. B}, 86:180503(R), Nov 2012.

\bibitem{Stanescu:PRL12}
Tudor~D. Stanescu, Sumanta Tewari, Jay~D. Sau, and S.~Das~Sarma.
\newblock To close or not to close: The fate of the superconducting gap across
  the topological quantum phase transition in majorana-carrying semiconductor
  nanowires.
\newblock {\em Phys. Rev. Lett.}, 109:266402, Dec 2012.

\bibitem{Rainis:PRB13}
Diego Rainis, Luka Trifunovic, Jelena Klinovaja, and Daniel Loss.
\newblock Towards a realistic transport modeling in a superconducting nanowire
  with majorana fermions.
\newblock {\em Phys. Rev. B}, 87:024515, Jan 2013.

\bibitem{Pientka:PRL12}
Falko Pientka, Graham Kells, Alessandro Romito, Piet~W. Brouwer, and Felix von
  Oppen.
\newblock Enhanced zero-bias majorana peak in the differential tunneling
  conductance of disordered multisubband quantum-wire/superconductor junctions.
\newblock {\em Phys. Rev. Lett.}, 109:227006, Nov 2012.

\bibitem{Bagrets:PRL12}
Dmitry Bagrets and Alexander Altland.
\newblock Class $d$ spectral peak in majorana quantum wires.
\newblock {\em Phys. Rev. Lett.}, 109:227005, Nov 2012.

\bibitem{Liu:PRL12}
Jie Liu, Andrew~C. Potter, K.~T. Law, and Patrick~A. Lee.
\newblock Zero-bias peaks in the tunneling conductance of spin-orbit-coupled
  superconducting wires with and without majorana end-states.
\newblock {\em Phys. Rev. Lett.}, 109:267002, Dec 2012.

\bibitem{Sau:13}
Jay~D. Sau and S.~Das Sarma.
\newblock Density of states of disordered topological
  superconductor-semiconductor hybrid nanowires.
\newblock {\em arxiv:1305.0554}, 2013.

\bibitem{Lee:PRL12}
Eduardo J.~H. Lee, Xiaocheng Jiang, Ram\'on Aguado, Georgios Katsaros,
  Charles~M. Lieber, and Silvano De~Franceschi.
\newblock Zero-bias anomaly in a nanowire quantum dot coupled to
  superconductors.
\newblock {\em Phys. Rev. Lett.}, 109:186802, Oct 2012.

\bibitem{Finck:PRL13}
A.~D.~K. Finck, D.~J. Van~Harlingen, P.~K. Mohseni, K.~Jung, and X.~Li.
\newblock Anomalous modulation of a zero-bias peak in a hybrid
  nanowire-superconductor device.
\newblock {\em Phys. Rev. Lett.}, 110:126406, Mar 2013.

\bibitem{Lee:13}
Eduardo J.~H. Lee, Xiaocheng Jiang, Manuel Houzet, Ramon Aguado, Charles~M.
  Lieber, and Silvano~De Franceschi.
\newblock Probing the spin texture of sub-gap states in hybrid
  superconductor-semiconductor nanostructures.
\newblock Preprint arXiv:1302.2611.

\bibitem{Churchill:PRB13}
H.~O.~H. Churchill, V.~Fatemi, K.~Grove-Rasmussen, M.~T. Deng, P.~Caroff, H.~Q.
  Xu, and C.~M. Marcus.
\newblock Superconductor-nanowire devices from tunneling to the multichannel
  regime: Zero-bias oscillations and magnetoconductance crossover.
\newblock {\em Phys. Rev. B}, 87:241401, Jun 2013.

\bibitem{Nayak:RMP08}
C.~Nayak, S.H. Simon, A.~Stern, M.~Freedman, and S.~Das~Sarma.
\newblock Non-abelian anyons and topological quantum computation.
\newblock {\em Rev. Mod. Phys.}, 80(3):1083--1159, 2008.

\bibitem{Kitaev:P01}
A~Yu Kitaev.
\newblock Unpaired majorana fermions in quantum wires.
\newblock {\em Phys. Usp.}, 44(10S):131, 2001.

\bibitem{Fu:PRB09}
Liang Fu and C.~L. Kane.
\newblock Josephson current and noise at a
  superconductor/quantum-spin-hall-insulator/superconductor junction.
\newblock {\em Phys. Rev. B}, 79(16):161408, Apr 2009.

\bibitem{Kwon:EPJB03}
H.J. Kwon, K.~Sengupta, and V.M. Yakovenko.
\newblock Fractional ac josephson effect in p-and d-wave superconductors.
\newblock {\em Eur. Phys. J. B}, 37(3):349--361, 2003.

\bibitem{Jiang:PRL11}
Liang Jiang, David Pekker, Jason Alicea, Gil Refael, Yuval Oreg, and Felix von
  Oppen.
\newblock Unconventional josephson signatures of majorana bound states.
\newblock {\em Phys. Rev. Lett.}, 107:236401, Nov 2011.

\bibitem{Dominguez:PRB12}
Fernando Dominguez, Fabian Hassler, and Gloria Platero.
\newblock Dynamical detection of majorana fermions in current-biased nanowires.
\newblock {\em Phys. Rev. B}, 86:140503, Oct 2012.

\bibitem{Rokhinson:NP12}
Leonid~P. Rokhinson, Xinyu Liu, and Jacek~K. Furdyna.
\newblock The fractional a.c. josephson effect in a
  semiconductor-superconductor nanowire as a signature of majorana particles.
\newblock {\em Nat Phys}, 8(11):795--799, 11 2012.

\bibitem{Sau:12b}
Jay~D. Sau, Erez Berg, and Bertrand~I. Halperin.
\newblock On the possibility of the fractional ac josephson effect in
  non-topological conventional superconductor-normal-superconductor junctions.
\newblock {\em arxiv:1206.4596}, 2012.

\bibitem{Badiane:PRL11}
Driss~M. Badiane, Manuel Houzet, and Julia~S. Meyer.
\newblock Nonequilibrium josephson effect through helical edge states.
\newblock {\em Phys. Rev. Lett.}, 107:177002, Oct 2011.

\bibitem{San-Jose:PRL12a}
Pablo San-Jose, Elsa Prada, and Ram\'on Aguado.
\newblock ac josephson effect in finite-length nanowire junctions with majorana
  modes.
\newblock {\em Phys. Rev. Lett.}, 108:257001, Jun 2012.

\bibitem{Pikulin:PRB12}
D.~I. Pikulin and Yuli~V. Nazarov.
\newblock Phenomenology and dynamics of a majorana josephson junction.
\newblock {\em Phys. Rev. B}, 86:140504, Oct 2012.

\bibitem{Doh:S05}
Yong-Joo Doh, Jorden~A. van Dam, Aarnoud~L. Roest, Erik P. A.~M. Bakkers,
  Leo~P. Kouwenhoven, and Silvano De~Franceschi.
\newblock Tunable supercurrent through semiconductor nanowires.
\newblock {\em Science}, 309(5732):272--275, 2005.

\bibitem{Nilsson:NL12}
H.~A. Nilsson, P.~Samuelsson, P.~Caroff, and H.~Q. Xu.
\newblock Supercurrent and multiple andreev reflections in an insb nanowire
  josephson junction.
\newblock {\em Nano Lett.}, 12(1):228--233, 2012.

\bibitem{Alicea:RPP12}
Jason Alicea.
\newblock New directions in the pursuit of majorana fermions in solid state
  systems.
\newblock {\em Rep. Prog. Phys.}, 75:076501, 02 2012.

\bibitem{Ghosh:PRB10}
Parag Ghosh, Jay~D. Sau, Sumanta Tewari, and S.~Das~Sarma.
\newblock Non-abelian topological order in noncentrosymmetric superconductors
  with broken time-reversal symmetry.
\newblock {\em Phys. Rev. B}, 82:184525, Nov 2010.

\bibitem{Cayao:UP}
Jorge Cayao, Pablo San-Jose, Elsa Prada, and Ramon Aguado.
\newblock To be published.

\bibitem{Cuevas:PRB96}
J.~C. Cuevas, A.~Mart\'in-Rodero, and A.~Levy Yeyati.
\newblock Hamiltonian approach to the transport properties of superconducting
  quantum point contacts.
\newblock {\em Phys. Rev. B}, 54:7366--7379, Sep 1996.

\bibitem{Beenakker:92}
CWJ Beenakker.
\newblock Three" universal" mesoscopic josephson effects.
\newblock In {\em Transport phenomena in mesoscopic systems: proceedings of the
  14th Taniguchi symposium, Shima, Japan, November 10-14, 1991}, page 235.
  Springer-Verlag, 1992.

\bibitem{Chevallier:PRB12}
D.~Chevallier, D.~Sticlet, P.~Simon, and C.~Bena.
\newblock Mutation of andreev into majorana bound states in long
  superconductor-normal and superconductor-normal-superconductor junctions.
\newblock {\em Phys. Rev. B}, 85:235307, Jun 2012.

\bibitem{Averin:PRL95}
D.~Averin and A.~Bardas.
\newblock ac josephson effect in a single quantum channel.
\newblock {\em Phys. Rev. Lett.}, 75:1831--1834, Aug 1995.

\bibitem{Sun:PRB02}
Qing-feng Sun, Hong Guo, and Jian Wang.
\newblock Hamiltonian approach to the ac josephson effect in
  superconducting-normal hybrid systems.
\newblock {\em Phys. Rev. B}, 65:075315, Jan 2002.

\bibitem{Rodrigo:PRL12}
J.~G. Rodrigo, V.~Crespo, H.~Suderow, S.~Vieira, and F.~Guinea.
\newblock Topological superconducting state of lead nanowires in an external
  magnetic field.
\newblock {\em Phys. Rev. Lett.}, 109:237003, Dec 2012.

\bibitem{Johansson:PRB99}
G.~Johansson, E.~N. Bratus, V.~S. Shumeiko, and G.~Wendin.
\newblock Resonant multiple andreev reflections in mesoscopic superconducting
  junctions.
\newblock {\em Phys. Rev. B}, 60:1382--1393, Jul 1999.

\bibitem{Yeyati:PRL03}
A.~Levy Yeyati, A.~Mart\'in-Rodero, and E.~Vecino.
\newblock Nonequilibrium dynamics of andreev states in the kondo regime.
\newblock {\em Phys. Rev. Lett.}, 91:266802, Dec 2003.

\bibitem{Jonckheere:PRB09}
T.~Jonckheere, A.~Zazunov, K.~V. Bayandin, V.~Shumeiko, and T.~Martin.
\newblock Nonequilibrium supercurrent through a quantum dot: Current harmonics
  and proximity effect due to a normal-metal lead.
\newblock {\em Phys. Rev. B}, 80:184510, Nov 2009.

\bibitem{Cuevas:PRB01}
J.~C. Cuevas and M.~Fogelstr\"om.
\newblock Quasiclassical description of transport through superconducting
  contacts.
\newblock {\em Phys. Rev. B}, 64:104502, Aug 2001.

\bibitem{Chauvin:PRL07}
M.~Chauvin, P.~vom Stein, D.~Esteve, C.~Urbina, J.~C. Cuevas, and A.~Levy
  Yeyati.
\newblock Crossover from josephson to multiple andreev reflection currents in
  atomic contacts.
\newblock {\em Phys. Rev. Lett.}, 99:067008, Aug 2007.

\bibitem{Cheng:PRB12}
Meng Cheng and Roman~M. Lutchyn.
\newblock Josephson current through a
  superconductor/semiconductor-nanowire/superconductor junction: Effects of
  strong spin-orbit coupling and zeeman splitting.
\newblock {\em Phys. Rev. B}, 86:134522, Oct 2012.

\bibitem{Lim:PRB12}
Jong~Soo Lim, Lloren{\c c} Serra, Rosa L\'opez, and Ram\'on Aguado.
\newblock Magnetic-field instability of majorana modes in multiband
  semiconductor wires.
\newblock {\em Phys. Rev. B}, 86:121103, Sep 2012.

\bibitem{Haug:07}
H.~Haug and A.P. Jauho.
\newblock {\em Quantum kinetics in transport and optics of semiconductors},
  volume 123.
\newblock Springer, 2007.

\end{thebibliography}


\end{document}